\documentclass[aps,twocolumn,groupedaddress]{revtex4-1}

\usepackage{amssymb, amsmath}
\usepackage{graphicx}
\usepackage{dcolumn}
\usepackage{bm}
\usepackage{longtable}
\usepackage{float}
\usepackage{multirow}
\begin{document}


\title{Mobility anisotropy of two-dimensional semiconductors}

\author{Haifeng Lang,$^{1,*}$ Shuqing Zhang,$^{2,}$}
\altaffiliation{These authors contributed equally to this work.}
\author{Zhirong Liu$^{1,2,3,}$}
\email{LiuZhiRong@pku.edu.cn}
\affiliation{$^1$College of Chemistry and Molecular Engineering, Peking University, Beijing 100871, China}
\affiliation{$^2$Center for Nanochemistry, Academy for Advanced Interdisciplinary Studies,
Peking University, Beijing 100871, China}
\affiliation{$^3$State Key Laboratory for Structural Chemistry of Unstable and Stable Species,
Beijing National Laboratory for Molecular Sciences, Peking University, Beijing 100871, China}

\begin{abstract}
The carrier mobility of anisotropic two-dimensional (2D) semiconductors under longitudinal acoustic (LA) phonon scattering
was theoretically studied with the deformation potential theory. Based on Boltzmann equation
with relaxation time approximation, an analytic formula of intrinsic anisotropic mobility was deduced,
which shows that the influence of effective mass to the mobility anisotropy is larger than that of
deformation potential constant and elastic modulus. Parameters were collected for various anisotropic 2D materials
(black phosphorus, Hittorf's phosphorus, BC$_2$N, MXene, TiS$_3$, GeCH$_3$) to calculate their mobility anisotropy.
It was revealed that the anisotropic ratio was overestimated in the past.
\end{abstract}

\maketitle
\section{INTRODUCTION}
The successful isolation of graphene in 2004\cite{Novoselov2004} led us into the brand new world of two-dimensional (2D)
materials.\cite{Geim2007,Deng2015} As the lecture title given by Richard P.~Feynman in 1959,\cite{Feynman1959}
``There's plenty of room at the bottom''.
Since graphene was born, unforeseen luxuriant physical and chemical properties of this atomically thin material
have attracted rising attention at extremely fast rate in the past years.\cite{Abergel2010,Allen2010}
For example, the unique ballistic transport and extraordinarily high carrier mobility greatly expanded
graphene's potential applications.\cite{Bolotin2008,Morozov2008} However, everything has its drawback.
The zero bandgap severely limits graphene's application in electronics.\cite{Meric2008}
Therefore, some efforts have been moved to explore the potentials of other 2D layered semiconductor
materials.\cite{Bhimanapati2015,Jinying2013} Representative systems include graphynes,\cite{Srinivasu2012,Li2010,Zhang2016}
transition metal dichalcogenides (TMDs),\cite{Cai2014,Zhang2014} black phosphorus (BP)\cite{Qiao2014,Xia2014}
and transition metal carbides and nitrides (MXenes)\cite{SR2016,Zhang2015}. They retain the one-atom-thin nature of graphene,
and provide applicable bandgaps, making them hopeful to be used in flexible electronics, photodetectors,
thin-film transistors and other devices.\cite{Bhimanapati2015,Xia2014}
On the other hand, suitable bandgap is necessary but not sufficient for a well-performing electronic component.
The carrier mobility is also crucial.\cite{Bhimanapati2015}

Some 2D materials are isotropic,\cite{Long2011,Xi2012,Cai2014,Wu2015} while others are
anisotropic.\cite{Qiao2014,Xia2014,SR2016,Zhang2015} For anisotropic 2D semiconductors,
their electrons and phonons have different behaviors along different directions in the plane,
leading to angle-dependent mechanical, optical and electrical response.
These unique properties may create unprecedented possibilities to design novel sensors with anisotropic
crystalline orientation, optical absorption and scattering, carrier mobility and electronic
conductance.\cite{Wu2015,AFM2016,Fei2014,NC2015}
Here, we focus on the theoretical study of the anisotropic carrier mobility.

Despite of the importance of carrier mobility, the theory of intrinsic mobility for anisotropic 2D semiconductors
was not well developed. For example, a widely adopted formula in the literature was given
as\cite{Qiao2014,SR2016,JPCL2014,Schusteritsch2016}
\begin{equation}
{{\mu}^{(\text{tr})}}=
\frac{e{{\hbar }^{3}}{{C}^{(\text{tr})}}}
{{{k}_{\text{B}}}T{{m}_{\text{d}}}{{m}^{(\text{tr})}}{{\left| E_{1}^{(\text{tr})} \right|}^{2}}},
\end{equation}
where the superscript \lq\lq(tr)\rq\rq was used to indicate that the corresponding quantities are defined
in the transport direction. $\mu$ is the carrier mobility. $e$ is the elementary charge,
$\hbar$ is the reduced Planck constant, ${k}_{\text{B}}$ is the Boltzmann's constant,
and $T$ is the temperature. ${m}^{\text{(tr)}}$ is the effective mass of charge carriers
(electrons and holes) along the transport direction (either $m_x$ or $m_y$ along the $x$ and $y$ directions, respectively),
and ${m}_{\text{d}}$ is the equivalent density-of-state mass defined as
${m}_{\text{d}}=\sqrt{{{m}_{x}}{{m}_{y}}}$. ${C}^{\text{(tr)}}$ is the 2D elastic modulus of the longitudinal strain
in the propagation directions, and ${E_1}^{\text{(tr)}}$ is the deformation potential constant
defined as the energy shift of the band edge position with respect to the strain. ${C}^{\text{(tr)}}$
and ${E_1}^{\text{(tr)}}$ in Eq.~(1) come from the influence of acoustic phonons. Eq.~(1) implies
that the mobility in a specified direction is determined only by $C$ and $E_1$ in the same direction
but is independent on those in the perpendicular direction. This is, however, logically incorrect,
because moving carriers would be inevitably scattered by phonons from all directions. In this study,
based on the Boltzmann equation, we will deduce an analytical formula of intrinsic mobility for anisotropic
semiconductors, where the mobility in one direction is indeed determined by $C$ and $E_1$ along all directions.

The rest of the paper is organized as follows. In Section II, the contributions of anisotropic $m$,
$C$ and $E_1$ on $\mu$ were theoretically studied separately. In Section III, the obtained formula
was applied to numerically analyze the mobility anisotropy of some 2D materials, which showed
that the anisotropy in most systems is weaker than what has been previously thought.
In Section IV, a historic perspective on the mobility of 2D electron gas (2DEG) was provided
to demonstrate its relation to that of 2D materials. Finally, we summarized our results and made some conclusions.

\section{Theoretical analyses on the anisotropic mobility}

\subsection{General consideration}

The carrier mobility of a sample is determined by various scattering processes.
A primary source of scatters for charge carriers is acoustic phonons, which cannot be removed at finite temperature
and thus determines the intrinsic mobility of the material. Most mobility predictions on 2D semiconductors
are based on the consideration of scattering by acoustic phonons. 
Theoretically, the intrinsic mobility caused by acoustical phonons can be described by the deformation potential
theory,\cite{Bardeen1950} where the atomic displacement associated with a long-wavelength acoustic phonon
leads to a deformation of the crystal, and in turn, to a shift of the band edge and scattering
between different eigen states.

In the spirit of the deformation potential theory, the shift of the band edge ($\Delta {{E}_{\text{edge}}}$)
is proportional to the longitudinal strain $\varepsilon (\mathbf{q})$
caused by longitudinal acoustic (LA) vibrational modes (phonons) with a wave vector $\mathbf{q}$:
\begin{equation}
\Delta {E_\text{edge}}={E_1}(\mathbf{q})\varepsilon(\mathbf{q}),
\label{Eq2}
\end{equation}
where the deformation potential constant $E_1$ depends on the direction of longitudinal strain and phonons
for anisotropic materials. The contribution of transverse acoustic (TA) phonons is ignored as in
the usual deformation potential theory. Under the Fermi-golden rule and the second quantization of the phonons,
the scattering probability of an electron from eigen state $\mathbf{k}$ to $\mathbf{k}'$
caused by LA phonons can be written as\cite{Xi2012,Price1981,Li2014}
\begin{equation}
{W_{\mathbf{k},\mathbf{{k}'}}}=\frac{2\pi{k_\text{B}}T{E_1} ( \mathbf{q})^{2}}{A\hbar C ( \mathbf{q})}
\delta ( {\varepsilon _\mathbf{k}}-{\varepsilon _{\mathbf{{k}'}}}),
\label{Eq3}
\end{equation}
where $A$ is the area of 2D sample and $C(\mathbf{q})$ is the elastic modulus caused by $\varepsilon(\mathbf{q})$.
The momentum conservation law requires that $\mathbf{q}=\mathbf{k}'-\mathbf{k}$.
Both the emission and the absorption of the phonons were considered in obtaining Eq.~(\ref{Eq3}),
and the temperature is much higher than the characteristic degenerate [Bloch-Gr{\"u}neisen (BG)] temperature.
The relaxation time for an electron in $\mathbf{k}$, denoted as $\tau(\mathbf{k})$, is thus given by
\begin{equation}
\frac{1}{\tau (\mathbf{k})}=\frac{A}{4\pi^2}\int{{W_{\mathbf{k},\mathbf{k}'}}
\left(1-\frac{{{\mathbf{v }}_{\mathbf{k}}}\cdot{{\mathbf{v}}_{{\mathbf{{k}'}}}}}
{{{ | {{\mathbf{v }}_{\mathbf{k}}} |}^2}}\right){{\text{d}}^2}\mathbf{k'}},
\label{Eq4}
\end{equation}
where $\mathbf{v}_{\mathbf{k}}$ is the group velocity.
The Boltzmann equation is the basis for the classical and semi-classical theories of transport processes.
It has been widely used in studying thermal, mass and  electrical conductivities under
weak driving forces. Based on the Boltzmann equation with the relaxation time approximation,
the 2D conductivity tensor is solved to be\cite{Xi2012,Li2014,Han2014}
\begin{equation}
\overset{\scriptscriptstyle\leftrightarrow}{\sigma }=2{e^2}\int{\tau (\mathbf{k})}
\frac{\partial n_{\text{F}} (\varepsilon_{\mathbf{k}}) }
{\partial \varepsilon_{\mathbf{k}} }
\mathbf{v }_{\mathbf{k}} \mathbf{v }_{\mathbf{k}}
\frac{{{\text{d}}^{2}}\mathbf{k}}{{{ ( 2\pi )}^2}},
\label{Eq5}
\end{equation}
where $\varepsilon_{\mathbf{k}}$ is the eigen energy of state $\mathbf{k}$,
and $n_\text{F}{(\varepsilon_{\mathbf{k}})}$ is the equilibrium Fermi-Dirac distribution.
The mobility along the $x$ direction is thus
\begin{equation}
{\mu _x}=\frac{{\sigma _{xx}}}{ne},
\label{Eq6}
\end{equation}
where $n=2\int{{n_{\text{F}}}({\varepsilon_\mathbf{k}})}\frac{{{\text{d}}^2}\mathbf{k}}{{{ ( 2\pi  )}^2}}$
is the carrier density.
Eqns. (\ref{Eq2}-\ref{Eq6}) provide the general framework to calculate the intrinsic mobility of 2D materials under LA phonons.
The mobility anisotropy may arise from anisotropic $\varepsilon_{\mathbf{k}}$
(which is related the anisotropic effective mass), $E_1(\mathbf{q})$ or $C(\mathbf{q})$,
which will be analyzed in details as follows.

\subsection{Anisotropic mass}

When only the effective mass is anisotropic, the energy dispersion is described as
\begin{equation}
{\varepsilon_\mathbf{k}}=\frac{{\hbar ^2}k_{x}^{2}}{2{m_x}}+\frac{{\hbar ^2}k_{y}^2}{2{m_y}},
\label{Eq7}
\end{equation}
where the $x$ and $y$ directions are chosen to be along the primary axes of the energy dispersion.
Making use of the coordinate transformation
\begin{equation}
\left\{ \begin{aligned}
{{{\tilde{k}}}_{x}}=\frac{{{k}_{x}}}{\sqrt{{{m}_{x}}}} \\
{{{\tilde{k}}}_{y}}=\frac{{{k}_{y}}}{\sqrt{{{m}_{y}}}}
\end{aligned}\right. ,
\label{Eq8}
\end{equation}
it is straight forward to derive from Eqns.~(\ref{Eq2}-\ref{Eq6}) to get the relaxation time
\begin{equation}
\tau (\mathbf{k})=\frac{{\hbar^3}{C_{11}}}{{k_{\text{B}}}TE_{1}^2\sqrt{{m_x}{m_y}}}
\label{Eq9}
\end{equation}
and the mobility
\begin{equation}
{\mu_x}=\frac{e{{\hbar }^3}{C_{11}}}{{k_\text{B}}TE_{1}^{2}{{\left( {m_x} \right)}^{\frac{3}{2}}}
{{\left( {{m}_{y}} \right)}^{\frac{1}{2}}}},
\label{Eq10}
\end{equation}
where ${E_1}\equiv {E_1}(\mathbf{q})$ and ${C_{11}}\equiv {C_{11}}(\mathbf{q})$ are isotropic.
$\tau(\mathbf{k})$ is independent on $\mathbf{k}$ even if the effective mass is anisotropic in this case.
Eq.~(\ref{Eq10}) is identical to Eq.~(1) if both ${C}^{\text{(tr)}}$ and ${E_1}^{\text{(tr)}}$ are isotropic,
i.e., Eq.~(1) is valid when only the effective mass is anisotropic.

\subsection{Elliptic deformation potential}

We now consider the case that only the deformation potential is anisotropic
while both effective mass and elastic modulus keep isotropic.
Strain is second-order tensor, so a longitudinal strain with any specified direction
can be decomposed into three components in 2D systems: two uniaxial strains (along $x$ and $y$ directions, respectively)
and a shear strain. If the system has mirror reflection symmetry, the contribution of the shear component to
the deformation potential disappears, and then $E_1(\mathbf{q})$ can be expressed as
\begin{equation}
{E_1}(\mathbf{q})={E_{1x}}{\cos^2}{{\theta}_{\mathbf{q}}}+{E_{1y}}{\sin^2}{\theta_{\mathbf{q}}},
\label{Eq11}
\end{equation}
where $\theta_{\mathbf{q}}$ is the polar angle of $\mathbf{q}$,
while $E_{1x}$ and $E_{1y}$ are deformation potential constants
along $x$ and $y$ directions, respectively.
Combined with Eqns.~(\ref{Eq3}, \ref{Eq11}), the integration in Eq.~(4) gives
\begin{equation}
\frac{1}{\tau(\mathbf{k})}=\frac{m{k_\text{B}}T}{{\hbar^3}{{C}_{11}}}\left[\bar{E}_{1}^{2}
+\frac{{{ (\Delta{E_1})}^2}}{2}-{{{\bar{E}}}_1}\Delta{E_1}\cos( 2{\theta_{\mathbf{k}}})\right],
\label{Eq12}
\end{equation}
where $\theta_{\mathbf{k}}$ is the polar angle of $\mathbf{k}$,
while $\bar{E}_{1}$ and $\Delta{E_1}$ are notations defined as
\begin{equation}
\left\{ \begin{aligned}
   {{{\bar{E}}}_1}=\frac{{E_{1y}}+{E_{1x}}}{2} \\
  \Delta {E_1}=\frac{{E_{1y}}-{E_{1x}}}{2}
\end{aligned}\right. .
\label{Eq13}
\end{equation}
$\tau(\mathbf{k})$ is anisotropic here, being distinct from the result of Eq.~(\ref{Eq9}) under anisotropic effective mass.
The mobility is obtained as
\begin{equation}
\begin{array}{l}
{{\mu }_{x}}\\
=\frac{ e\int{\frac{{{\hbar }^{3}}{{C}_{11}}}{m{{k}_{\text{B}}}T}}\cdot \frac{{{k}^{2}}{{\cos }^{2}}{{\theta }_{\mathbf{k}}}}
{\bar{E}_{1}^{2}+\frac{{{\left( \Delta {{E}_{1}} \right)}^{2}}}{2}
-{{{\bar{E}}}_{1}}\Delta {{E}_{1}}\cos \left( 2{{\theta }_{\mathbf{k}}} \right)}\cdot
\frac{{{\hbar }^{2}}}{{{m}^{2}}}\cdot \frac{\partial {{n}_{\text{F}}}({{\varepsilon }_{\mathbf{k}}})}
{\partial {{\varepsilon }_{\mathbf{k}}}}{{\text{d}}^{2}}\mathbf{k}}
{\int{\frac{\partial {{n}_{\text{F}}}({{\varepsilon }_{\mathbf{k}}})}{\partial {{\varepsilon }_{\mathbf{k}}}}}
{{\text{d}}^{2}}\mathbf{k}}  \\
=
\text{ }\frac{e{{\hbar }^{3}}{{C}_{11}}}{{{m}^{2}}{{k}_{\text{B}}}T}
\left( \frac{A+B-\sqrt{{{A}^{2}}-{{B}^{2}}}}{B\sqrt{{{A}^{2}}-{{B}^{2}}}} \right)
\end{array}
\label{Eq14}
\end{equation}
with the notations
\begin{equation}
\left\{ \begin{aligned}
  & A=\bar{E}_{1}^{2}+\frac{{{\left( \Delta {{E}_{1}} \right)}^{2}}}{2} \\
 & B={{{\bar{E}}}_{1}}\Delta {{E}_{1}} \\
\end{aligned} \right.
\label{Eq15}
\end{equation}
Eq.~(\ref{Eq14}) is a bit complicated. To see the anisotropic effect more clearly, we rewrite it into
\begin{equation}
{\mu_x}=\frac{e{\hbar}^3{C_{11}}}{{m^2}{k_\text{B}}T{{\bar{E}_1}^2}}
\times \frac{1}{f\left( \frac{\Delta {E_1}}{{{\bar{E}}_{1}}} \right)},
\label{Eq16}
\end{equation}
where $f\left( \frac{\Delta {E_1}}{{{\bar{E}}_1}} \right)$ is a corrected factor due to the anisotropic effect:
\begin{equation}
f\left(\frac{\Delta{E_1}}{{{\bar{E}}_1}}\right)
=\frac{1}{\bar{E}_{1}^{2}}\cdot \frac{B\sqrt{{A^2}-{B^2}}}{A+B-\sqrt{{A^2}-{B^2}}}.
\label{Eq17}
\end{equation}
The curve of $f\left( \frac{\Delta {E_1}}{{{\bar{E}}_1}} \right)$ is given in Fig.~1.
For $\Delta {E_1}=0$, it gives $f=1$, consistent with the isotropic result.
Within the examined range, $f\left( \frac{\Delta {E_1}}{{{\bar{E}}_1}} \right)$
can be well reproduced by a quadratic function:
\begin{equation}
f\left(\frac{\Delta{E_1}}{\bar{E}_1}\right)
=1-0.5\frac{\Delta{E_1}}{\bar{E}_1}+0.3{\left (\frac{\Delta{E_1}}{\bar{E}_1}\right)}^2
\label{Eq18}
\end{equation}
as demonstrated as the solid line in Fig.~1.
With the quadratic approximation, the mobility under anisotropic deformation potential is simplified into
\begin{equation}
{\mu_x}=\frac{e{{\hbar }^{3}}{{C}_{11}}}
{{{m}^{2}}{{k}_{\text{B}}}T\left( \frac{9E_{1x}^{2}+7{{E}_{1x}}{{E}_{1y}}+4E_{1y}^{2}}{20} \right)}.
\label{Eq19}
\end{equation}
It can be seen that the mobility along the $x$ direction not only depends on the deformation potential
along the same direction $(E_{1x})$, but also depends on that along its perpendicular direction $(E_{1y})$.

\begin{figure}
\includegraphics[width =0.45\textwidth]{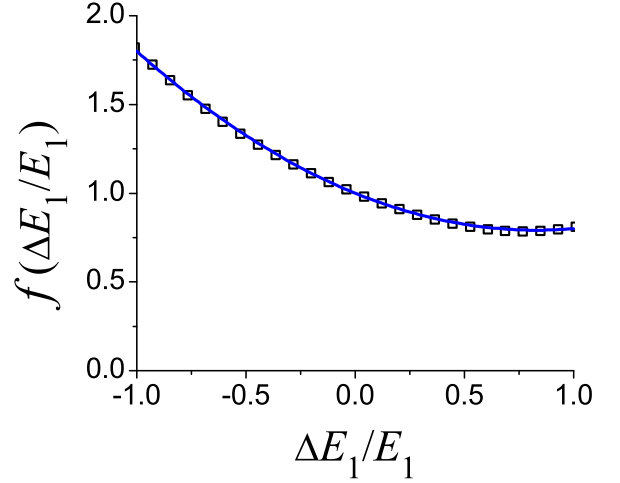}
\caption{
The corrected factor $f\left(\frac{\Delta{E_1}}{{{\bar{E}}_1}}\right)$ due to the anisotropic deformation potential.
Scattering points were calculated with Eq.~(\ref{Eq17}),
while solid line is a quadratic approximation as given in Eq.~(\ref{Eq18}).
}
\end{figure}

\subsection{Elliptic elastic constant}

For the anisotropic effect of elastic modulus $C(\mathbf{q})$,
usually only the values along the primary axes (taken as $x$ and $y$ directions) were calculated in the literature.
As an approximation, we express $C(\mathbf{q})$ as:
\begin{equation}
C(\mathbf{q})={C_{11}}{\cos^2}{\theta_\mathbf{q}}+{C_{22}}{\sin^2}{\theta_\mathbf{q}},
\label{Eq20}
\end{equation}
where $C_{11}$ and $C_{22}$ are 2D elastic constants along $x$ and $y$ directions, respectively.
The effective mass and deformation potential are kept isotropic.
The relaxation time is obtained as:
\begin{equation}
\frac{1}{\tau (\mathbf{k})}=
\frac{m{{k}_{\text{B}}}TE_{1}^{2}}
{{{\hbar }^{3}}}\left[ \frac{1+\frac{{\bar{C}}}{\Delta C}\cos ( 2{{\theta }_{\mathbf{k}}})}
{\sqrt{{{{\bar{C}}}^{2}}-{{\left( \Delta C \right)}^{2}}}}-\frac{\cos ( 2{{\theta }_{\mathbf{k}}})}{\Delta C} \right],
\label{Eq21}
\end{equation}
where
\begin{equation}
\left\{ \begin{aligned}
 \bar{C}=\frac{{{C}_{11}}+{{C}_{22}}}{2} \\
 \Delta C=\frac{{{C}_{22}}-{{C}_{11}}}{2}
\end{aligned}\right. .
\label{Eq22}
\end{equation}
The mobility is given as
\begin{equation}
\begin{array}{l}
{{\mu }_{x}} \\
 =\frac{\int{\frac{e{{\hbar }^{5}}}
{m^3 {{k}_{\text{B}}}TE_{1}^{2}}} \cdot
\frac{{{k}^{2}}{{\cos }^{2}}{{\theta }_{\mathbf{k}}}}
{\frac{1}{\sqrt{{{{\bar{C}}}^{2}}-{{\left( \Delta C \right)}^{2}}}}
-\frac{{\bar{C}}}{\Delta C}
\left( \frac{1}{{\bar{C}}}-\frac{1}{\sqrt{{{{\bar{C}}}^{2}}-{{\left( \Delta C \right)}^{2}}}} \right)
\cos \left( 2{{\theta }_{\mathbf{k}}} \right)}
\cdot \frac{\partial {{n}_{\text{F}}}({{\varepsilon }_{\mathbf{k}}})}
{\partial {{\varepsilon }_{\mathbf{k}}}}{{\text{d}}^{2}}\mathbf{k}}
{\int{\frac{\partial {{n}_{\text{F}}}({{\varepsilon }_{\mathbf{k}}})}
{\partial {{\varepsilon }_{\mathbf{k}}}}}{{\text{d}}^{2}}\mathbf{k}}  \\
=\text{ }\frac{e{{\hbar }^{3}}}{{{m}^{2}}{{k}_{\text{B}}}TE_{1}^{2}}
\left( \frac{I+J-\sqrt{{{I}^{2}}-{{J}^{2}}}}{J\sqrt{{{I}^{2}}-{{J}^{2}}}} \right),
\end{array}
\label{Eq23}
\end{equation}

where
\begin{equation}
\left\{ \begin{array}{l}
  I=\frac{1}{\sqrt{{{{\bar{C}}}^{2}}-{{\left( \Delta C \right)}^{2}}}} \\
 J=\frac{{\bar{C}}}{\Delta C}\left( \frac{1}{{\bar{C}}}-\frac{1}
{\sqrt{{{{\bar{C}}}^{2}}-{{\left( \Delta C \right)}^{2}}}} \right)
\end{array}\right.
\label{Eq24}
\end{equation}
Eq.~(\ref{Eq23}) can be expanded to the linear order of $\Delta C$ to give a simplified result:
\begin{equation}
{\mu_x}=\frac{e{{\hbar }^3}}{{m^2}
{k_\text{B}}TE_{1}^{2}}\left( \frac{5{C_{11}}+3{C_{22}}}{8} \right),
\label{Eq25}
\end{equation}
which shows that the mobility along the $x$ direction depends on both the elastic constants along $x$ and $y$ directions.

\section{Results and Discussion}

\begin{figure}
\includegraphics[width =0.45\textwidth]{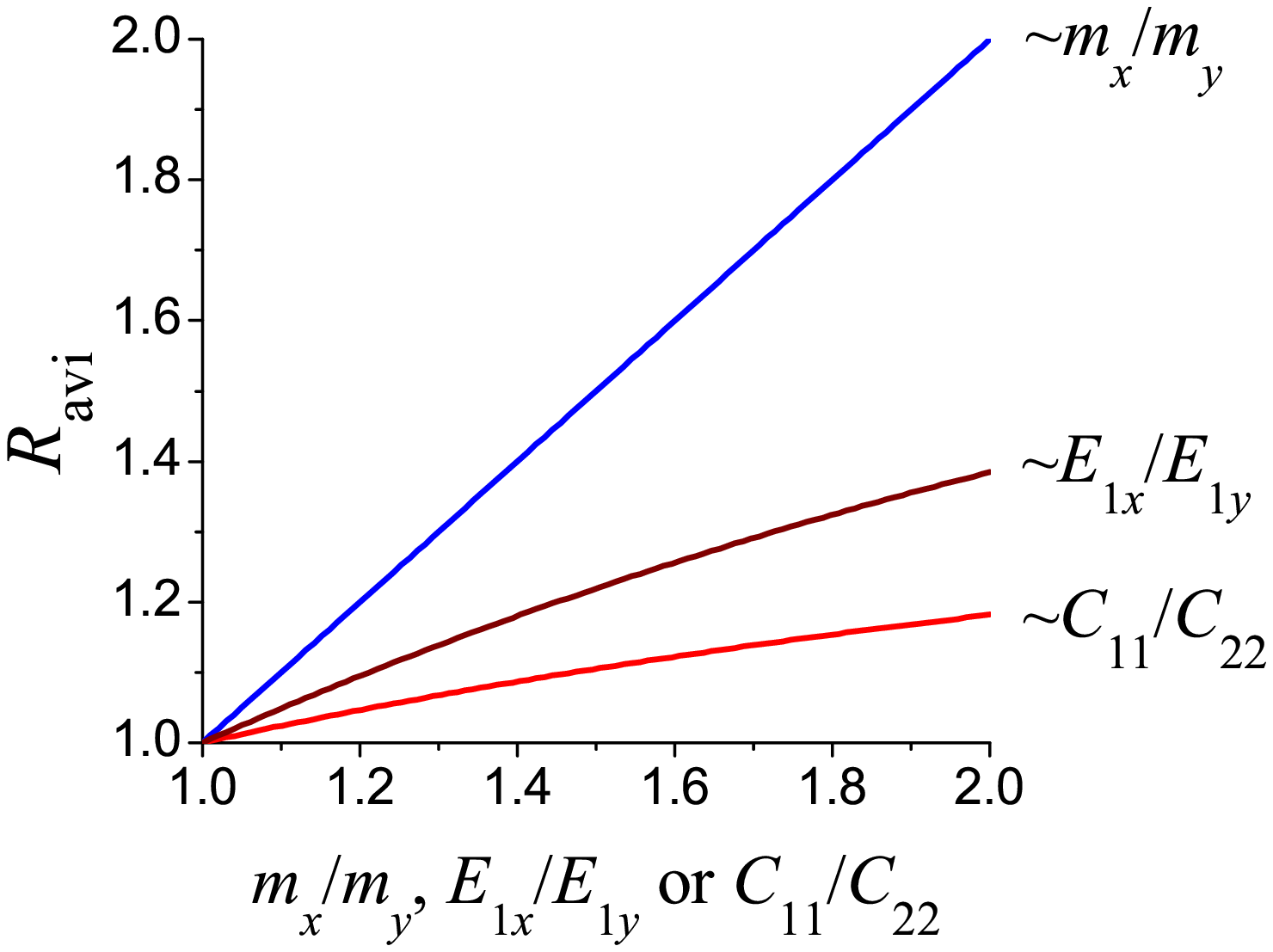}
\caption{
Anisotropic ratio of mobility $(R_\text{ani})$ as functions of $\frac{m_x}{m_y}$,
$\frac{E_{1x}}{E_{1y}}$ and $\frac{C_{11}}{C_{22}}$. Results are calculated with Eq.~(\ref{Eq27}).
Except the parameter being varied in each line, ${\frac{m_x}{m_y}}=1$, ${\frac{E_{1x}}{E_{1y}}}=1$,
$\frac{C_{11}}{C_{22}}=1$.
}
\end{figure}

\subsection{Combined anisotropic effects on mobility}

In the section above, the anisotropic effects of the effective mass,
the deformation potential and the elastic modulus on the mobility of 2D semiconductors were analyzed separately
to give analytical results. When all these anisotropic factors appear together in a system,
it is too complicated to achieve an analytical solution. Therefore, we propose to express the mobility approximately
by combining different anisotropic factors directly:
\begin{widetext}
\begin{equation}
{{\mu }_x}=\frac{e{\hbar^3}}{{k_\text{B}}T{{\left( {{m}_{x}} \right)}^{\frac{3}{2}}}
{{\left( {{m}_{y}} \right)}^{\frac{1}{2}}}}\left( \frac{A+B-\sqrt{{A^2}-{B^2}}}
{B\sqrt{{A^2}-{B^2}}} \right)\left( \frac{I+J-\sqrt{{I^2}-{J^2}}}{J\sqrt{{I^2}-{J^2}}} \right),
\label{Eq26}
\end{equation}
\end{widetext}
where $A$ and $B$ are functions of deformation potential whose definition was given in Eq.~(\ref{Eq15}),
while $I$ and $J$ are functions of elastic modulus, whose definition was given in Eq.~(\ref{Eq24}).
With the low order approximation, a concise form is achieved as:
\begin{equation}
{{\mu }_{x}}=\text{ }\frac{e{{\hbar }^{3}}\left( \frac{5{{C}_{11}}+3{{C}_{22}}}{8} \right)}
{{{k}_{\text{B}}}T{{\left( {{m}_{x}} \right)}^{\frac{3}{2}}}
{{\left( {{m}_{y}} \right)}^{\frac{1}{2}}}\left( \frac{9E_{1x}^{2}+7{{E}_{1x}}{{E}_{1y}}+4E_{1y}^{2}}{20} \right)}.
\label{Eq27}
\end{equation}
Anisotropic effective mass is the main contributor to the mobility anisotropy.
To measure the mobility anisotropy, we define an anisotropic ratio $(R_\text{ani})$ as:
\begin{equation}
{R_\text{ani}}=\frac{\max ({\mu_x},{\mu_y})}{\min ({\mu_x},{\mu_y})},
\label{Eq28}
\end{equation}
which is equal to 1.0 for isotropic systems and is larger than 1.0 for anisotropic systems.
The variation of $R_\text{ani}$ with various parameters are demonstrated in Fig.~2.
For anisotropic mass acting along with ${\frac{m_x}{m_y}}=2$, it yields $R_\text{ani}=2.0$.
In comparison, it is only $R_\text{ani}=1.18$ and
$R_\text{ani}=1.38$ for ${\frac{C_{11}}{C_{22}}}=2$ and ${\frac{E_{1x}}{E_{1y}}}=2$, respectively.
Therefore, the anisotropy contribution from elastic constant and deformation potential is much weaker than
that from the energy dispersion (effective mass).
Consistently, for materials with Dirac cone and zero bandgap, the anisotropic contribution is also dominated
by the energy dispersion (Fermi velocity) while the contribution from deformation potential is nearly zero.\cite{Li2014,Deng2015,Zeren2015}

\subsection{Numerical results}

\begin{table*}
\caption{Predicted mobility anisotropy of some representative 2D semiconductors.}
\begin{ruledtabular}
\begin{tabular}{ccccccccccc|ccc|ccc}
\multicolumn{2}{c}{\multirow{2}{*}{System}} &
\multicolumn{1}{c}{\multirow{2}{*}{$m_x$}} & \multicolumn{1}{c}{\multirow{2}{*}{$m_y$}} &
\multicolumn{1}{c}{\multirow{2}{*}{$E_{1x}$}} & \multicolumn{1}{c}{\multirow{2}{*}{$E_{1y}$}} &
\multicolumn{1}{c}{\multirow{2}{*}{$C_{11}$}} & \multicolumn{1}{c}{\multirow{2}{*}{$C_{22}$}} &
\multicolumn{3}{c}{old} &\multicolumn{3}{c}{new} &\multicolumn{3}{c}{simplified} \\
\cline{9-17}
 & & & & & & & &$\mu_x$ &$\mu_y$ &$R_\text{ani}$ &$\mu_x$ &$\mu_y$ &$R_\text{ani}$ &$\mu_x$ &$\mu_y$ &$R_\text{ani}$\\
\hline
\multirow{2}{*}{BP\cite{Qiao2014}}
 &e &0.17 &1.12&2.72&7.11&28.9&102&1.12&0.08&14.0&0.69&0.09&7.40&0.80&0.40&7.64\\
   &h&0.15&6.35&2.5&0.15&28.9&102&0.67&16.0&23.9&2.37&0.16&14.6&2.77&0.18&15.1\\
\hline
\multirow{2}{*}{2-BP\cite{Qiao2014}}
&e&0.18&1.13&5.02&7.35&57.5&195&0.60&0.15&4.00&0.81&0.14&5.58&0.70&0.13&5.76\\
&h&0.15&1.81&2.45&1.63&57.5&195&2.70&1.80&1.50&6.40&0.85&7.28&5.53&0.76&7.52\\
\hline
\multirow{2}{*}{3-BP\cite{Qiao2014}}
&e&0.16&1.15&5.85&7.63&85.9&287&0.78&0.21&3.71&1.17&0.19&6.06&1.01&0.17&6.25\\
&h&0.15&1.12&2.49&2.24&85.9&287&4.80&2.70&1.78&9.72&1.80&5.24&8.41&1.61&5.40\\
\hline
\multirow{2}{*}{4-BP\cite{Qiao2014}}
&e&0.16&1.16&5.92&7.58&115&379&1.02&0.28&3.64&1.54&0.25&6.08&1.33&0.22&6.26\\
&h&0.14&0.97&3.16&2.79&115&379&4.80&2.90&1.66&9.66&1.94&4.83&8.38&1.74&4.97\\
\hline
\multirow{2}{*}{5-BP\cite{Qiao2014}}
&e&0.15&1.18&5.79&7.53&146&480&1.47&0.38&3.87&2.19&0.32&6.66&1.91&0.29&6.86\\
&h&0.14&0.89&3.40&2.97&146&480&5.90&3.80&1.55&11.1&2.45&4.42&9.68&2.19&4.55\\
\hline
\multirow{2}{*}{HP\cite{Schusteritsch2016}}
 &e &0.69&3.58&1.40&0.66&49.7&49.9&0.50&0.43&1.16&0.76&0.21&3.65&0.76&0.21&3.65\\
 &h&1.24&2.45&1.26&0.18&49.7&49.9&0.31&7.68&24.8&0.61&0.60&1.01&0.62&0.61&1.01\\
\hline
\multirow{2}{*}{BC$_2$N\cite{JPCL2014}}
&e&0.15&0.41&1.87&4.25&307&400&52.5&3.70&14.2&22.3&5.59&3.98&22.1&5.56&3.97\\
        &h&0.16&2.22&2.13&4.33&307&400&14.8&0.27&54.9&7.66&0.40&19.3&7.62&0.39&19.3\\
\hline
\multirow{2}{*}{2-BC$_2$N\cite{JPCL2014}}
&e&0.16&0.40&1.86&4.13&771&769&118&9.61&12.3&52.9&14.6&3.62&52.2&14.7&3.61\\
&h&0.18&0.58&2.15&4.21&771&769&60.5&4.95&12.2&32.0&7.24&4.43&32.2&7.27&4.43\\
\hline
\multirow{2}{*}{3-BC$_2$N\cite{JPCL2014}}
&e&0.17&0.41&0.79&2.79&1023&901&809&23.4&34.5&177&42.3&4.22&179&42.5&4.20\\
&h&0.20&0.66&3.41&2.82&1023&901&27.0&10.3&2.62&28.1&9.06&3.10&28.0&9.05&3.10\\
\hline
\multirow{2}{*}{4-BC$_2$N\cite{JPCL2014}}
&e&0.17&0.42&0.95&3.30&1254&1285&651&22.4&29.1&161&39.1&4.14&163&39.3&4.12\\
&h&0.21&0.87&2.80&3.47&1254&1285&37.7&6.15&6.13&32.1&7.01&4.58&32.1&7.02&4.58\\
\hline
\multirow{2}{*}{5-BC$_2$N\cite{JPCL2014}}
&e&0.18&0.43&2.0&0.88&1856&1571&200&364&1.81&289&169&1.71&290&170&1.71\\
&h&0.23&1.0&3.44&2.63&1856&1571&31.3&10.2&3.06&34.2&8.61&3.97&34.1&8.59&3.97\\
\hline
\multirow{2}{*}{Ti$_2$CO$_2$\cite{SR2016}}
 &e&0.38&3.03&9.17&4.71&253&256&0.15&0.07&2.08&0.23&0.04&5.83&0.23&0.04&5.84\\
            &h&0.09&0.13&3.25&5.28&253&256&50.1&12.8&3.91&34.1&17.9&1.90&34.1&18.0&1.90\\
\hline
\multirow{2}{*}{Ti$_2$CO$_2$\cite{Zhang2015}}
&e&0.44&4.53&5.71&0.85&267&265&0.61&0.25&2.41&0.56&0.11&5.31&0.55&0.10&5.34\\
            &h&0.14&0.16&1.66&2.60&267&265&74.1&22.5&3.29&66.1&46.4&1.43&65.9&46.3&1.42\\
\hline
\multirow{2}{*}{TiS$_3$\cite{Dai2015}}
&e&1.47&0.41&0.73&0.94&81.3&145&1.01&13.9&13.7&2.89&10.6&3.66&2.99&10.9&3.64\\
       &h&0.32&0.98&3.05&-3.8&81.3&145&1.12&0.15&8.07&2.30&0.71&3.26&4.16&1.12&3.73\\
\hline
\multirow{2}{*}{GeCH$_3$\cite{Jing2015}}
&e&0.03&0.19&12.7&12.5&51.7&49.6&6.71&0.12&53.7&3.71&0.51&7.31&3.72&0.51&7.31\\
  &h&0.04&0.31&6.24&6.28&51.7&49.6&14.0&0.19&75.3&7.07&0.83&8.56&7.07&0.83&8.56\\
  \hline
\multirow{3}{*}{Sc$_2$CF$_2$\cite{NS2016}}
 &e&0.25&1.46&2.26&1.98&193&182&5.03&1.07&4.70&5.62&1.02&5.48&5.62&1.02&5.48\\
&h(u)&2.25&0.44&1.91&-4.7&193&182&0.48&0.39&1.25&0.61&1.20&2.43&0.42&1.02&1.96\\
&h(l)&0.46&2.65&-5.0&2.2&193&182&0.31&0.26&1.18&0.94&0.41&2.93&0.78&0.26&2.32\\
\hline
\multirow{3}{*}{Sc$_2$C(OH)$_2$\cite{NS2016}}
&e&0.50&0.49&-2.7&-2.6&173&172&2.06&2.19&1.06&2.18&2.22&1.02&2.18&2.22&1.02\\
&h(u)&5.01&0.27&-3.5&-9.9&173&172&0.05&0.11&2.24&0.02&0.20&11.7&0.02&0.20&11.7\\
&h(l)&0.29&1.91&-10&-3.2&173&172&0.16&0.24&1.45&0.29&0.07&4.05&0.29&0.07&4.07\\
\hline
\multirow{3}{*}{Hf$_2$CO$_2$\cite{SR2016}}
&e&0.23&2.16&10.6&7.10&294&291&0.33&0.08&4.27&0.44&0.06&7.72&0.44&0.06&7.72\\
&h(u)&0.42&0.16&7.64&2.30&294&291&0.92&26.0&28.1&1.67&7.13&4.28&1.68&7.21&4.26\\
&h(l)&0.16&0.41&2.02&7.42&294&291&34.3&1.00&34.3&8.04&1.86&4.33&8.14&1.88&4.31\\
\hline
\multirow{3}{*}{Zr$_2$CO$_2$\cite{SR2016}}
&e&0.27&1.87&13.9&5.21&265&262&0.15&0.15&1.02&0.26&0.06&4.59&0.26&0.06&4.60\\
&h(u)&0.16&0.38&9.84&1.80&265&262&1.37&17.5&12.8&2.72&2.20&1.24&2.74&2.21&1.24\\
&h(l)&0.36&0.16&5.45&6.04&265&262&2.08&3.71&1.78&1.98&4.17&2.10&1.98&4.17&2.10\\
\end{tabular}
\end{ruledtabular}
\begin{flushleft}
`e' and `h' denote `electron' and `hole', respectively.
$m_x$ and $m_y$ are measured as the ratio with $m_0$ (the electron mass in vacuum).
$E_{1x}$ and $E_{1y}$ are in units of eV.
$C_{11}$ and $C_{22}$ are in units of J/m$^2$.
$\mu_x$ and $\mu_y$ are in units of 10$^3$ cm$^2$V$^{-1}$s$^{-1}$.
The values of $\mu_x$, $\mu_y$, $E_{1x}$, $E_{1y}$,  $C_{11}$ and $C_{22}$ are extracted from references as indicated.
$\mu_x$ and $\mu_y$ are calculated in three ways: (old) same as in original references
(largely based Eq.~(1)), (new) Eq.~(\ref{Eq26}), and (simplified) Eq.~(\ref{Eq27}).
The anisotropic ratio $R_\text{ani}$ is calculated by Eq.~(\ref{Eq28}).``upper'' and ``lower'' sub-bands in the literature are represented by (u) and (l) here.  For few layer samples, for $n$ layer sample, which is expressed as $n-$sample type, such as $n$-BP and $n$-BC$_2$N.
\end{flushleft}
\end{table*}

To numerically evaluate the mobility anisotropy of 2D semiconductors and examine
how the new formula [Eqns.~(\ref{Eq26}, \ref{Eq27})] produce results different from the old one [Eq.~(1)],
data for various anisotropic materials were collected from the literature, including BP,\cite{Qiao2014}
single-layer Hittorf's phosphorus (HP),\cite{Schusteritsch2016} BC$_2$N,\cite{JPCL2014}
TiS$_3$,\cite{Dai2015,Aierken2016} GeCH$_3$,\cite{Jing2015} Ti$_2$CO$_2$,\cite{SR2016,Zhang2015}
Hf$_2$CO$_2$,\cite{SR2016} Zr$_2$CO$_2$,\cite{SR2016} Sc$_2$CF$_2$ and Sc$_2$C(OH)$_2$.\cite{NS2016}
Analysis results for some representative systems are listed in Table I. For these systems, Eq.~(\ref{Eq26}) and Eq.~(\ref{Eq27}) give very close results,
suggesting the simplified Eq.~(\ref{Eq27}) is a good approximation to the full form of Eq.~(\ref{Eq26}).
However, the difference between new and old methods is distinct, as discussed below.

Undoped BP is $p$-type semiconductor. Related experiments on thin-layer BP suggested that
the hole mobility were larger than electron one and the mobility along the $x$ (armchair) direction
were greater than that along the $y$ direction.\cite{Xia2014,Morita1986,Li2015,Mishchenko2015}
However, the calculation of single-layer BP by the old formula gave opposite results
of ${\mu_x(\text{h})}<{\mu_x(\text{e})}$ and ${\mu_x(\text{h})}<{\mu_y(\text{h})}$,
as shown in Table I. Instead, under the same parameters,
the new formula produces results with the trend consistent with the experiments.
The origin of the discrepancy between the old and new formula comes from the fact
that the old formula overestimates the contribution of the deformation potential to the mobility anisotropy.
It predicted that the anisotropy ratio is proportional to ${\left(\frac{E_{1y}}{E_{1x}}\right)}^{2}$[see Eq.~(1)].
Since single-layer BP has $E_{1x}$ (= 2.5 eV) much larger than $E_{1y}$ (= 0.15 eV) for holes,
it predicted ${\mu_x}<{\mu_y}$. However, according to the new formula,
the contribution of the deformation potential to the mobility anisotropy is actually weak,
where the main contributor is the effective mass. $m_x$ of BP is smaller than $m_y$, so ${\mu_x}>{\mu_y}$
under the new formula, which is consistent with the prediction by Kubo-Nakano-Mori method based on electron-phonon scattering matrices\cite{Rudenko2016} and charged-impurity scattering theory.\cite{Yue2016} Moreover, they are in agreement with the experimental observations.\cite{Xia2014,Li2015,Mishchenko2015}

TiS$_3$ monolayer is a new 2D material predicted to possess novel electronic properties.\cite{Dai2015,Aierken2016,Island2015}
First-principles calculations showed that TiS$_3$ is a direct-gap semiconductor with a bandgap of 1.02 eV,
close to that of bulk silicon.\cite{Dai2015} With the old method, TiS$_3$ was predicted to possess high mobility up to
$14\times10^3~\text{cm}^2 \text{V}^{-1} \text{s}^{-1}$ for electrons in the $y$ direction $[\mu_y(\text{e})]$,
and more remarkably, the mobility is highly anisotropic, i.e., $\mu_y(\text{e})$ is about 14 times higher
than $\mu_x(\text{e})$ and is even two orders of magnitude higher than $\mu_y(\text{h})$.\cite{Dai2015}
With the new method, however, the obtained anisotropy is much smaller. The re-calculated $\mu_y(\text{e})$
is $10.6\times10^3~\text{cm}^2 \text{V}^{-1} \text{s}^{-1}$, close to the old value, but the re-calculated
$\mu_x(\text{e})$ increases from $1.01\times10^3~\text{cm}^2 \text{V}^{-1} \text{s}^{-1}$
to $2.89\times10^3~\text{cm}^2 \text{V}^{-1} \text{s}^{-1}$, giving an anisotropic ratio of
only $R_\text{ani}=3.7$ (see Table I). The recalculated electrons/holes mobility ratio is 15 instead of 100,
suggesting that the potential in electron/hole separation is not so remarkable as previously thought.

\begin{figure}
\includegraphics[width =0.45\textwidth]{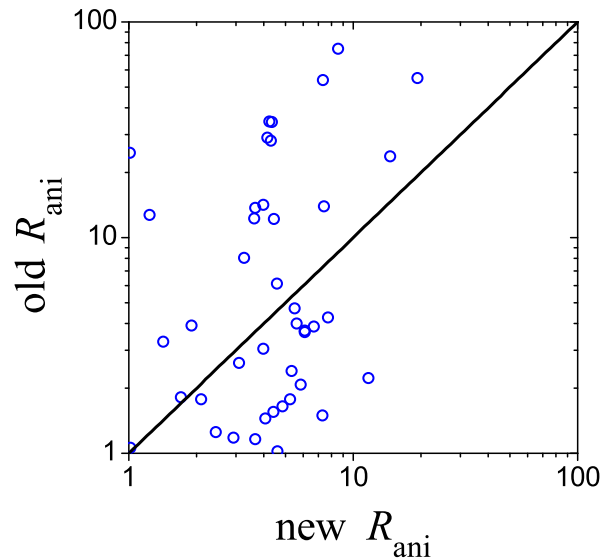}
\caption{Anisotropic ratio of mobility for various systems calculated by old and new methods,
i.e., Eq.~(1) and Eq.~(\ref{Eq26}), respectively.
Detailed data are provided in Table I.}
\end{figure}

The calculated anisotropy ratios from new and old methods are analyzed in Fig.~3.
The discrepancy is large, indicating that the old method is highly unreliable.
Overall, the old method is more likely to predict high anisotropy.
For example, among all 42 datapoints, only three were predicted by the new method to possess $R_\text{ani}>10$:
holes of BP (14.6), holes of BC$_2$N (19.3) and holes of Sc$_2$C(OH)$_2$ (11.7).
In comparison, the old method predicted 15 datapoints to have $R_\text{ani}>10$,
three of which possess $R_\text{ani}>50$: holes of BC$_2$N (54.9), and holes (75.3)
and electrons (53.7) of GeCH$_3$.

\section{HISTORICAL PERSPECTIVE}

To understand the state of art of mobility calculation in anisotropic 2D semiconductors by the deformation potential theory,
it is necessary to know the historical development. In this section, we make a brief survey on it,
and recognize some improper ways in the literature in calculating mobility.

The deformation potential theory was first proposed by Bardeen and Shockley in 1950 for three-dimension (3D)
non-polar semiconductors.\cite{Bardeen1950} With the development of metal-oxide-semiconductor
field-effect transistor (MOSFET) in the next years, scientists found that the electrons move
in the semiconductor-oxide interface of MOSFET, being free in 2D but tightly confined in the third dimension,
which could be described as a 2D sheet embedded in a 3D world. All the constructs with similar characteristics
were known as 2D electron gas (2DEG).\cite{Kawaji1969,Sato1971} In 1969, Kamaji extended
the deformation potential theory to the phonon-limited carrier mobility of 2DEG
in a semiconductor inversion layer by an inverted triangular well potential model, and a simple formula
was reported to calculate the lattice-scattering mobility of 2DEG:\cite{Kawaji1969}
\begin{equation}
\mu =\frac{e{{\hbar }^{3}}{{\rho }^{(3\text{D})}}\upsilon _{l}^{2}}{{{m}^{2}}{{k}_{\text{B}}}TE_{1}^{2}}
\cdot {{W}_{\text{eff}}},
\label{Eq29}
\end{equation}
where $\rho^{(3\text{D})}$ is the 3D mass density of the crystal,
$\upsilon_l$ is the velocity of longitudinal wave and $\rho{\upsilon_l}$ can be replaced by
3D elastic constant $C_{11}^{(3\text{D})}$. $W_\text{eff}$ is the effective thickness of the inversion layer
with a complex expression determined by dielectric constant of the material as well as the impurity
and free electron densities.\cite{Bardeen1950,Kawaji1969}
Then in 1981, Price applied the theory in a semiconductor hetero-layer to calculate the lattice-scattering
mobility,\cite{Price1981} where he described the layer for active carriers in 2DEG as square wells,
and obtained a simple expression for $W_\text{eff}$:
\begin{equation}
W_\text{eff}=\frac{2}{3}L,
\label{Eq30}
\end{equation}
where $L$ is the width of the square well.\cite{Price1981}
In anisotropic systems, effective mass $m$ and deformation potential constant $E_1$ are second-order tensors,
while elastic modulus $C$ is a fourth-order tensor, components of these tensors are not independent.\cite{Hinckley1990}
The mobility anisotropy of 2DEG on oxidized silicon surfaces could be attributed to the difference
in the effective mass and it was interpreted by Sat\^{o} {\it et al.}~in 1971 based on
an ellipsoidal constant-energy surface.\cite{Sato1971} With the anisotropic mass,
the mobility of 2DEG in inversion layer was modified into\cite{Kazuo1989,Takagi1994}
\begin{equation}
{{\mu }_{x}}=\frac{e{{\hbar }^{3}}{{\rho }^{(3\text{D})}}\upsilon _{l}^{2}}
{{{m}_{x}}{{m}_{\text{d}}}{{k}_{\text{B}}}TE_{1}^{2}}\cdot {{W}_{\text{eff}}},
\label{Eq31}
\end{equation}
where ${m_\text{d}}=\sqrt{{m_x}{m_y}}$.

2DEG in inversion layer is not real 2D system in the sense that it is always embedded in 3D material.
That is the reason why 3D parameter $\rho^{(3\text{D})}$ appeared in Eqns.~(\ref{Eq29}, \ref{Eq31}).
Graphene and other 2D crystals studied in recent years, on the other hand,
are real 2D systems since they could exist independently. As an important property,
their mobility attracted a lot of interest.\cite{Zhang2014,Qiao2014,Long2011,Northrup2011,Bruzzone2011}
Almost all of mobility calculations were based on the generation of Eqns. (29, 31) of 2DEG.
To generate the formula to real 2D systems, some ones assumed
$\rho^{(\text{3D})}W_\text{eff}=\rho^{(\text{2D})}$ to give\cite{Xi2012,Takagi1994,Northrup2011,Takagi1996}
\begin{equation}
{{\mu }_{x}}=\frac{e{{\hbar }^{3}}{{\rho }^{(2\text{D})}}\upsilon _{l}^{2}}{{{m}_{x}}
{{m}_{\text{d}}}{{k}_{\text{B}}}TE_{1}^{2}}=\frac{e{{\hbar }^{3}}C_{11}^{(2\text{D})}}
{{{m}_{x}}{{m}_{\text{d}}}{{k}_{\text{B}}}TE_{1}^{2}},
\label{Eq32}
\end{equation}
while some others assumed $\rho^{(\text{3D})}L=\rho^{(\text{2D})}$
to give\cite{Zhang2015,Dai2015,Aierken2016,Jing2015}
\begin{equation}
{{\mu }_{x}}=\frac{2e{{\hbar }^{3}}{{\rho }^{(2\text{D})}}\upsilon _{l}^{2}}{3{{m}_{x}}
{{m}_{\text{d}}}{{k}_{\text{B}}}TE_{1}^{2}}=\frac{2e{{\hbar }^{3}}C_{11}^{(2\text{D})}}
{3{{m}_{x}}{{m}_{\text{d}}}{{k}_{\text{B}}}TE_{1}^{2}}.
\label{Eq33}
\end{equation}
The generations were somehow arbitrary without necessary theoretical deduction.
For example, the factor 2/3 comes from 2DEG being confined in square well,
but the behaviors of electrons in real 2D systems have nothing to do with square well.
By comparing with Eq.~(\ref{Eq10}) we deduced above, it is recognized that
Eq.~(\ref{Eq32}) is valid when both deformation potential and elastic modulus are isotropic,
while Eq.~(\ref{Eq33}) is always improper. Another improper generation in the literature lay in the anisotropic effects.
Eq.~(\ref{Eq32}) was originally used to investigate the mobility of isotropic system
such as 2D hexagonal BN,\cite{Bruzzone2011} but it was later adopted
to study anisotropic systems such as BP.\cite{Qiao2014,Fei2014}
As we have revealed in the above sections, Eq.~(\ref{Eq32}) is actually not applicable
under anisotropic deformation potential and elastic modulus.

\section{Summary}

In summary, we have theoretically studied the LA-phonon-limited mobility for anisotropic 2D semiconductors
under the framework of the deformation potential theory. The influences of anisotropic deformation potential constant
and elastic modulus were analytically derived. It was shown that the mobility in one direction depends not only
on the parameters (effective mass, deformation potential constant and elastic modulus) along the same direction,
but also depends on those along its perpendicular direction. The mobility anisotropy is mainly contributed
by the anisotropic effective mass, while the distribution from the deformation potential constant and elastic modulus
is much weaker. Parameters for various anisotropic 2D materials were collected to calculate the mobility anisotropy.
It was demonstrated that the old formulas widely adopted in the literature were unreliable, and they were more likely
to overestimate the anisotropic ratio.

\begin{acknowledgments}

The authors thank Zhenzhu Li, Ting Cheng and Mei Zhou for helpful discussions.
This work was supported by the National Natural Science Foundation of China (Grant No. 21373015).

\end{acknowledgments}

\providecommand{\noopsort}[1]{}\providecommand{\singleletter}[1]{#1}%
%


\end{document}